\documentclass{article}
\usepackage{frascatiphys}
\usepackage{graphicx}
\usepackage{amsmath}
\usepackage{txfonts}

\usepackage{natbib}

\newcommand{\diffd}{\mathrm{d}}	

\newcommand{\apj}[1]{ApJ}

\begin{document}
\title{ 
Constraining Dark Matter models with extremely distant galaxies
}

 \author{
Marco Castellano        \\
{\em INAF - Osservatorio Astronomico di Roma} \\
Nicola Menci       \\
{\em INAF - Osservatorio Astronomico di Roma}\\
Andrea Grazian \\
{\em INAF - Osservatorio Astronomico di Roma}\\
Alexander Merle\\
{\em Max-Plank-Institut f\"ur Physik (Werner-Heisenberg-Institut)}\\
Norma G. Sanchez\\
{\em CNRS LERMA, Observatoire de Paris PSL, Sorbonne Universit\'es}\\
Aurel Schneider\\
{\em Institute for Astronomy, Department of Physics, ETH Zurich}\\
Maximilian Totzauer\\ 
{\em Max-Plank-Institut f\"ur Physik (Werner-Heisenberg-Institut)}
}
\maketitle
\baselineskip=11.6pt
\begin{abstract}
The investigation of distant galaxy formation and evolution is a powerful tool to constrain dark matter scenarios, supporting and in some cases surpassing other astrophysical and experimental probes. The recent completion of the Hubble Frontier Field (HFF) programme combining ultra-deep Hubble Space Telescope observations and the magnification power of gravitational lensing produced by foreground galaxy clusters has enabled the detection of the faintest primordial galaxies ever studied. Here we show how the number density of such primordial galaxies allows to constrain a variety of DM models alternative to CDM. In particular, it provides stringent limits on the mass of thermal WDM candidates, on the parameter space of sterile neutrino production models, and on other DM scenarios featuring particles in the keV mass range which is also supported by recent detections of a 3.5keV X-ray line. These constraints are robust and independent of the baryonic physics modeling of galaxy formation and evolution. Fuzzy DM (ultralight DM particles) results strongly disfavored.
\end{abstract}
\baselineskip=14pt

\section{Introduction}

Understanding the nature of the Dark Matter (DM) component of the Universe constitutes a key issue in fundamental physics and in cosmology. During the last two decades, investigations of the formation and growth of cosmic structures have progressively led to the adoption of the Cold Dark Matter (CDM) paradigm, where DM particles are characterized by thermal velocities small enough to produce negligible free streaming on the scales relevant to structure formation ~\citep[e.g.][]{Peebles:1982ff}. Typically, this corresponds either to assuming DM particles to be massive ($m_X>0.1$~GeV) or to be constituted by condensates of light axions (with mass $\sim 10^{-5}-10^{-1}$~eV). However, as of now, both direct~\citep[see, e.g.,][]{Aprile:2012nq} and indirect~\citep[see, e.g.,][]{Ackermann:2015zua} CDM detection experiments have failed to provide a definite confirmation of such a scenario. On the structure formation side, several critical issues are affecting the CDM scenario at the mass scales of dwarf galaxies ($M\approx 10^7-10^9$ $M_\odot$). These are all connected to the excess of power in the CDM power spectrum at such scales compared to a variety of observations.

The combination of astrophysical issues with the lack of detection of candidate particles has stimulated the interest toward different DM scenarios, characterized by power spectra with suppressed amplitude at small mass scales ($M\lesssim 10^8-10^9$ $M_\odot$) with respect to the CDM case. In particular, great attention has been given to Warm Dark Matter (WDM) scenarios, which assume DM to be composed by particles with masses $m_X$ in the~keV range that potentially provide a Dark Matter interpretation of the  claimed detection of an X-ray line in stacked observations of galaxy clusters and in the Perseus cluster~\citep{Bulbul:2014sua,Boyarsky:2014jta}. While WDM candidates may result from the freeze-out of particles initially in thermal equilibrium in the early Universe~\citep[like, e.g., gravitinos, see][for a review]{steffen2006}, a similar suppression at these scales can be obtained by a variety of models featuring particles in the~keV mass range with \emph{non-thermal} spectra, like sterile neutrinos. Finally, another proposed solution to the small-scale problems in galaxy formation is based on Bose condensates of ultra-light (pseudo) scalar field DM with mass $m_\psi\approx 10^{-22}$~eV~, often referred to as "Fuzzy" DM.

Existing astrophysical bounds on the thermal relic mass $m_X$, have been set with a variety of techniques \citep[e.g.][]{Polisensky:2010rw,Schultz2014}, the tightest constraints achieved so far being the $m_X\geq 3$ keV, derived by comparing small scale structure in the Lyman-$\alpha$ forest of high- resolution ($z > 4$) quasar spectra with hydrodynamical N-body simulations \citep[][]{Viel2013}. 

The abundance of low-mass cosmic structures provides an important key to constrain DM scenarios. In this context, the Hubble Frontier Field (HFF) programme has recently provided important information through the detection of ultra-faint, lensed galaxies at very high-redshifts. In fact, estimates of the UV luminosity function down to unprecedented faint magnitudes $M_{\rm UV}=-12.5$ at $z=6$ in~\citep{Livermore:2016mbs}, can be used to derive limits on the total number density of galaxies at early epochs.

In the present paper we summarise the results presented in \cite{Menci2016,Menci2017} where stringent constraints on DM models with suppressed power spectra by have been derived by comparing the maximum number density of DM halos $\overline{\phi}$ expected at redshift $z=6$ to the observed number density ${\phi}_{obs}$ of galaxies at the same redshift in the HFF.  The condition that observed galaxies cannot outnumber their host DM halos ($\overline{\phi}\geq {\phi}_{obs}$) directly leads to constraints on the set of parameters admitted for each DM model. {\it Remarkably, this technique provides a conservative approach which is not affected by uncertainties in the baryonic physics}, at variance with most of previous investigations of DM scenarios alternative to CDM.

\section{The halo mass function in dark-matter models with suppressed power spectra}

\subsection{Warm Dark Matter thermal relics}

The simplest alternative to CDM is provided by Warm Dark Matter models assuming DM to be the result from the freeze-out of particles  with mass in the keV range initially in thermal equilibrium in the early Universe. In these models, the population of low-mass galaxies is characterized by lower abundances and shallower central density profiles compared to Cold Dark Matter (CDM) due to the dissipation of small-scale density perturbations produced by the free-streaming of the lighter and faster DM particles. In this case, the mass of the DM particle completely determines the suppression of the density power spectrum compared to the CDM case 
 
The computation of the halo mass function for the WDM scenario is based on the standard procedure described and tested against N-body simulations in, e.g., \citet{Schneider2013, Angulo2013}.  
The differential halo mass function (per unit $log\,M$) based on the extended Press \& Schechter approach \citep[e.g.][]{Bond1991} reads:
\begin{equation}
 {d\,\phi\over d\,logM}={1\over 6}\,{\overline{\rho}\over M}\,f(\nu)\,{d\,log\,\sigma^2\over d\,log r}\,.
\end{equation}
Here $\nu\equiv \delta_c^2(t)/\sigma^2$ depends on the linearly extrapolated density for collapse in the spherical model $\delta_c=1.686/D(t)$ and $D(t)$ is the growth factor of DM perturbations. A spherical collapse model for which $f(\nu)=\sqrt{2\nu/\pi}\,exp(-\nu/2)$ is assumed.

The key quantity entering Eq. 1 is the variance of the linear power spectrum $P(k)$ of DM perturbations (in terms of the wave-number $k=2\pi/r$). Its dependence on the spatial scale $r$ of perturbations is:
\begin{equation} 
{d\,log\,\sigma^2\over d\,log\,r}=-{1\over 2\,\pi^2\,\sigma^2(r)}\,{P(1/r)\over r^3}.
\end{equation}

In  WDM scenarios the spectrum $P_{WDM}$ is suppressed with respect to the CDM case $P_{CDM}$ below a characteristic scale depending on the mass  $m_X$ of the WDM particles. In the case of relic thermalized particles, the suppression factor can be parametrized as \citep{Bode2001}:
\begin{equation}
{P_{WDM}(k)\over P_{CDM}(k)}=\Big[1+(\alpha\,k)^{2\,\mu}\Big]^{-10/\mu}\,.
\end{equation}
where $\mu=1.12$ and the quantity $\alpha$ is linked to the WDM free-streaming scale:
\begin{equation}
\alpha=0.049 \,
\Bigg[{\Omega_X\over 0.25}\Bigg]^{0.11}\,
\Bigg[{m_X\over {\rm keV}}\Bigg]^{-1.11}\,
\Bigg[{h\over 0.7}\Bigg]^{1.22}\,{h^{-1}\over \rm Mpc},  
\end{equation}
where $m_X$ is the WDM particle mass, $\Omega_X$ is the WDM density parameter ($\Omega_X$) and $h$ the Hubble constant in units of 100 km/s/Mpc.

The mass function is computed through Eq. 1 after substituting Eq. 2, with a power spectrum $P(k)=P_{WDM}(k)$ 
determined by the WDM particle mass $m_X$ after Eqs. 3 and 4. 

\subsection{Sterile neutrinos}

\subsubsection{Resonant production from mixing with active neutrinos}

A suppression to the power spectrum similar to the WDM case can be obtained by a variety of models featuring particles in the~keV mass range with \emph{non-thermal} spectra, like sterile neutrinos, the main difference being that in the case of non-thermal spectra, the production mechanism is essential in determining the suppression with respect to CDM. The minimal setup for sterile neutrino DM is the production via mixing with one or several active neutrino flavors. Active neutrinos are weakly interacting and are therefore in thermal equilibrium with other Standard Model particles in the early Universe. During that epoch, the sterile neutrino abundance builds up gradually via occasional oscillations from the active to the sterile sector. Combined limits from structure formation and X-ray observations \citep[e.g.][]{Merle:2015vzu} have recently ruled out \emph{non-resonant} production \citep[Dodelson-Widrow (DW) mechanism,][]{Dodelson:1993je}.

However, active-sterile oscillation may be enhanced by a resonance~\citep[Shi \& Fuller (SF) or \emph{resonant} production mechanism, RP,][]{Shi:1998km}, provided there exists a significant lepton asymmetry $L$ in the early Universe. Such a resonance allows for significantly smaller mixing angles $\theta$, relaxing the tight limits from X-ray observations. In this scenario, for any given sterile neutrino mass, the mixing angle is related to the adopted lepton asymmetry $L$, such that the parameter space of RP sterile neutrino models can be described in terms of combinations of sterile neutrino masses $m_{\nu}$ and mixing amplitudes $sin^2(2\theta)$. Each one of such combinations corresponds to a different momentum distribution, which strongly differs from a generic Fermi-Dirac form~\citep{Abazajian:2001nj}.

\subsubsection{Production from scalar decay}

Production from scalar decay (SD) is described by a generic model that invokes one real scalar singlet $S$ and (at least) one sterile neutrino $N$ beyond the Standard Model. The scalar singlet couples to the SM Higgs doublet $\Phi$ via a \emph{Higgs portal}, while the interaction between the scalar and the sterile neutrino is described by a Yukawa-type coupling.

The free parameters of the scalar decay model are: 1) the Higgs portal coupling $\lambda$, which determines the production rate and the kinematics of the scalar from the SM degrees of freedom of the Higgs doublet; 2) the Yukawa coupling $y$, which enters into the decay rate of the scalar and hence controls how fast the scalar decays into sterile neutrinos; 3) the mass of the scalar singlet, $m_S$, which determines which channels contribute to the production of scalars and thereby finally to the abundance of sterile neutrinos (see Sect. 2 of \citealt{Konig:2016dzg}); 4) the mass of the sterile neutrino $m_\nu$, strongly influencing the effects on cosmological structure formation. 

In \citet{Menci2017}, we treated $\lambda$, $y$, and $m_S$ as free parameters. For each triple of $\left(\lambda, y, m_S\right)$, we fixed the mass of the sterile neutrino by requiring it to reproduce the observed relic DM abundance. The interplay between the Higgs portal and the Yukawa coupling results in two different regimes: 1) for small Higgs portal couplings, the scalar itself is produced by freeze-in and is always strongly suppressed compared to its would-be equilibrium abundance. In this case, the relic abundance of sterile neutrinos (and hence the mass $m_\nu$) are independent of the Yukawa coupling $y$ for a fixed pair  $\left(m_S,\lambda\right)$. 2) When $\lambda$ is large enough to equilibrate the scalars, they will be subject to the well-known dynamics of freeze-out. In this regime, sterile neutrinos can be produced from scalars in equilibrium and from those decaying after freeze-out. Accordingly, the number density of steriles and thereby their mass $m_\nu$ can strongly depend on $y$ even for fixed $\left(m_S,\lambda\right)$.

\subsubsection{The halo MF of sterile neutrino models}
An  approach similar to the one adopted for thermal WDM is used for the sterile neutrino RP and SD models, but in this case the power spectrum is computed directly by solving the Boltzmann equation after computing the distribution function for all points of the parameter space. The resulting differential mass functions are characterized by a maximum value at masses close to the ``half-mode'' mass~\citep[e.g.,][]{Schneider:2011yu,Angulo2013},  the mass scale at which the spectrum is suppressed by 1/2 compared to CDM. This function depends strongly on the sterile neutrino mass; for RP models it also depends on the lepton asymmetry assumed and, hence, on the resulting mixing angle $\theta$; typical power spectra in such models yield half-mode masses ranging from $M_{hm}\approx 10^{10}\,M_\odot$ to $M_{hm}\approx 10^{8}\,M_\odot$. Correspondingly, the cumulative mass functions saturate to a maximum value  $\overline{\phi}(z)\approx \phi(M_{hm},z)$, defining the maximum number density of DM halos associated to the considered power spectrum. 

\subsection{Fuzzy Dark Matter}

Fuzzy DM models assume the DM to be composed of a non-relativistic Bose-Einstein condensate, so that the uncertainty principle counters gravity below a Jeans scale corresponding to the de Broglie wavelength of the ground state. In this case, the  suppression of  small scale structures and the formation of galactic cores in dwarf galaxies is in fact entirely due to the uncertainty principle, which counteracts gravity below the Jeans scale, corresponding to a mass scale $M_J=10^{7}\,M_{\odot}\,m_{22}^{-3/2}$~\citep{Marsh:2013ywa}, where $m_{22}\equiv m_{\psi}/10^{-22}$ eV. In such models, the DM mass $m_{\psi}$ ultimately determines all the relevant DM physical scales in structure formation, since it determines the scale below which an increase in momentum opposes any attempt to confine the particle any further. 

In the Fuzzy DM case, dedicated $N$-body simulations~\citep{Schive:2015kza} yield for the differential mass function the form 
\begin{equation}
 {\diffd\,\phi\over \diffd (\ln M)}={\diffd \,\phi\over \diffd (\ln M)}\bigg|_{\rm CDM} \cdot \Bigg[1+\bigg({M\over M_0}\bigg)^{-1.1}\Bigg]^{-1.2},
  \label{eq:HaloMassDerivativeFuzzy}
 \end{equation}
where $\left |\diffd \,\phi/ \diffd \left(\ln M\right)\right|_{\rm CDM}$ is the halo mass function in the CDM scenario. The auxiliary mass scale $M_0 = 1.6 \times 10^{10}\, (m_\psi/10^{-22}\,{\rm~eV})^{-4/3}\, M_\odot$, determining the  suppression of the halo mass function compared to the CDM case, depends on the Fuzzy DM candidate mass, and it plays a role analogous to the half-mode mass scale for sterile neutrino models.

\section{The observed galaxy number density at z$\sim$6}

The above mentioned halo number densities are compared to the observed number density ${\phi}_{obs}$ of galaxies derived by integrating the galaxy luminosity function (LF) at $z=6$ by ~\citet{Livermore:2016mbs} down to the faintest bin $M_{\rm UV}= -12.5$. Constraints on DM models are simply put by requiring that observed galaxies cannot outnumber their host DM halos ($\overline{\phi}\geq {\phi}_{obs}$). The reference luminosity function has been estimated from objects in the Abell~2744 and MACS~0416 cluster fields, selected on the basis of their photometric redshift. The UV LF with the corresponding 1-$\sigma$ uncertainties in each magnitude bin is estimated on the basis of the median magnification for each galaxy in the sample and is reported in Fig.~10 of~\citet{Livermore:2016mbs}. From this we have derived the observed cumulative number density $\phi_{obs}$ (and its confidence levels) through a Monte Carlo procedure. We extracted random  values $\Phi_{random}(M_{UV})$ of the luminosity function in each magnitude bin according to a Gaussian distribution with variance given by the relevant error bar.  Thus, for each simulation we produced a new realization of the luminosity function at $z=6$. From this, a cumulative number density $\phi_{random}$ has been derived by summing up the values of $\Phi_{random}(M_{UV})$ in all the observed magnitude bins in the range  $-22.5\leq M_{UV}\leq -12.5$.  We carried out $N_{sim}=10^7$ simulations to compute the probability distribution function (PDF) of the cumulative number density $\phi_{random}$. We obtain a median value  $log\,{\phi}_{obs}/{\rm Mpc}^{-3}=0.54$, while from the relevant percentiles of the PDF we  derive lower bounds  0.26, 0.01, and -0.32 at  1, 2, and 3-$\sigma$ confidence levels, respectively.  

\section{Results}

 \subsection{Thermal WDM}
 \begin{figure}[htb]
    \begin{center}
        {\includegraphics[scale=0.35]{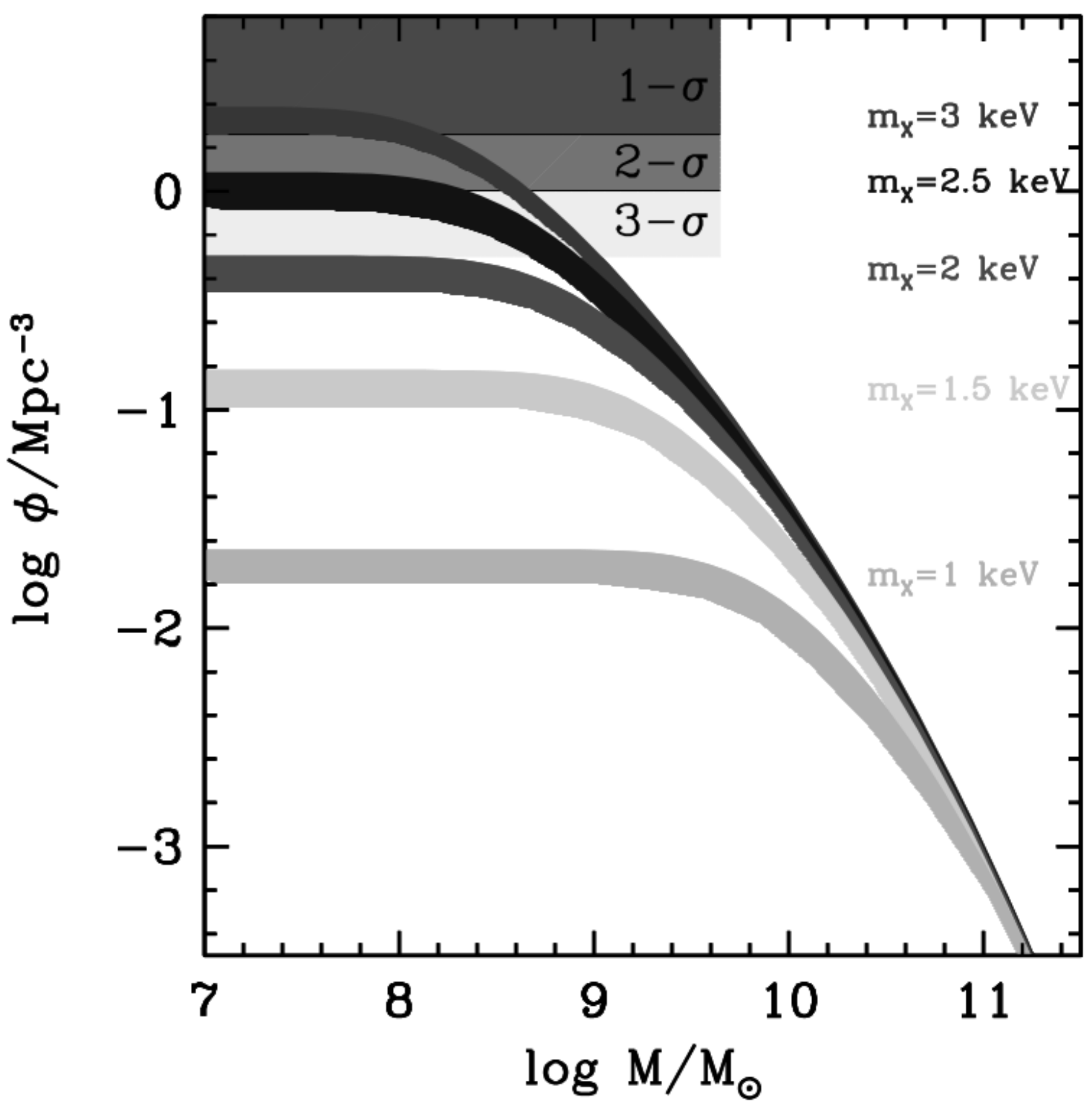}}\hspace{0.5cm}
        \caption{\it Adapted from \citet{Menci2016}: the cumulative mass functions computed at $z=6$ for different values of the WDM particle mass $m_X$ from 1 to 3 keV (bottom to top). The shaded areas correspond to the observed number density of HFF galaxies within 1-$\sigma$, 2-$\sigma$, and  3-$\sigma$  confidence levels.}
\label{fig1}
    \end{center}
\end{figure}

In Fig.~\ref{fig1} we show the cumulative mass function $\phi(>M)$ at $z=6$ for different assumed WDM particle masses. 
All the mass functions saturate to a maximum number density $\overline\phi_{m_X}\approx \phi (M_{hm})$. This is compared with the observed number density $\phi_{obs}$ of galaxies with $M_{UV}\leq -12.5$.  The condition $\phi_{obs}\leq \overline\phi_{m_X}$ yields  $m_X\gtrsim 2.9$ keV at 1-$\sigma$ level,  $m_X\geq 2.4$ keV at 2-$\sigma$ level, and $m_X\geq 2.1$ keV at 3-$\sigma$ level.

 \subsection{Sterile Neutrino from resonant productions}

\begin{figure}[htb]
    \begin{center}
        {\includegraphics[scale=0.25]{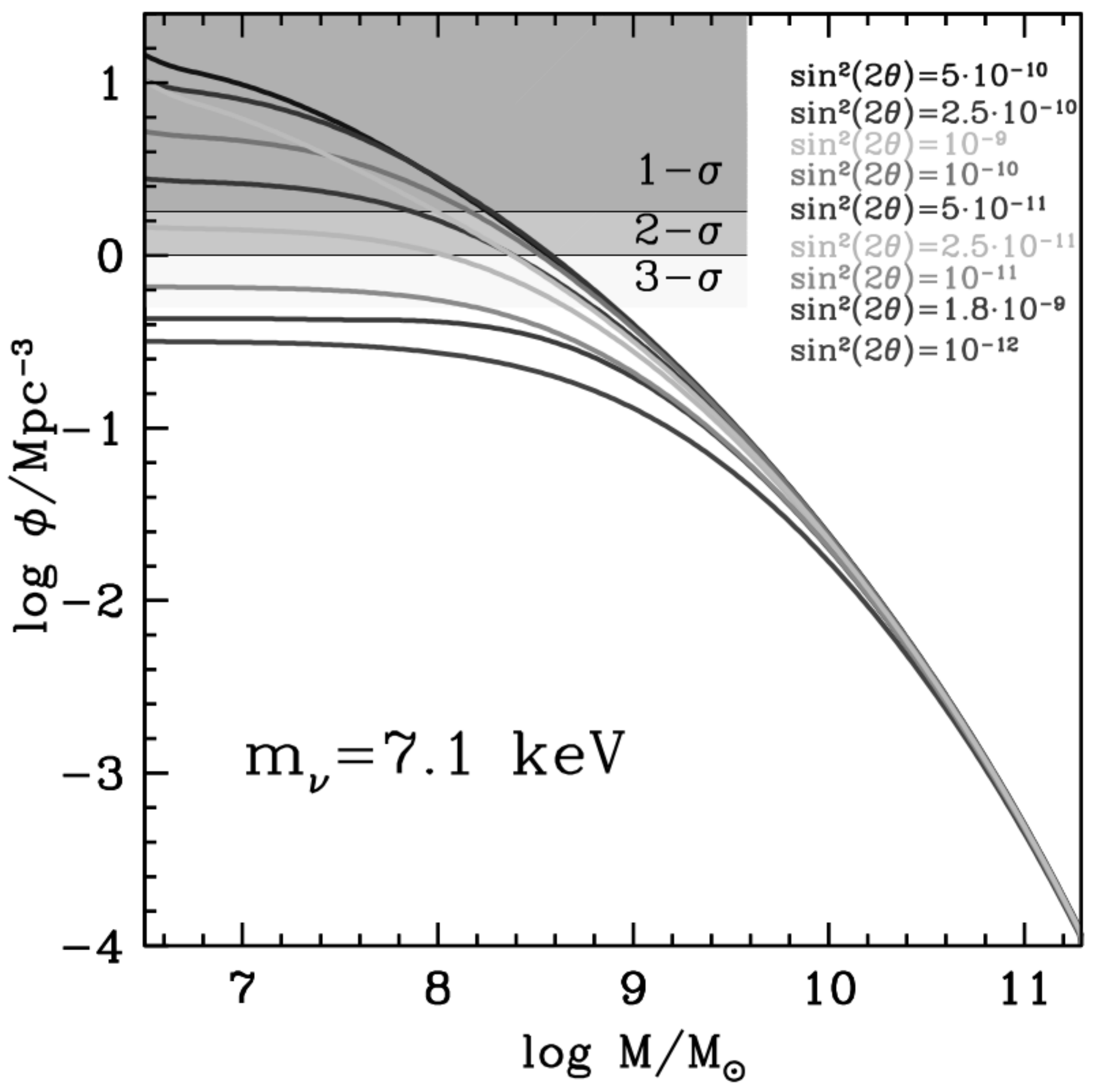}}\hspace{0.5cm}
        {\includegraphics[scale=0.25]{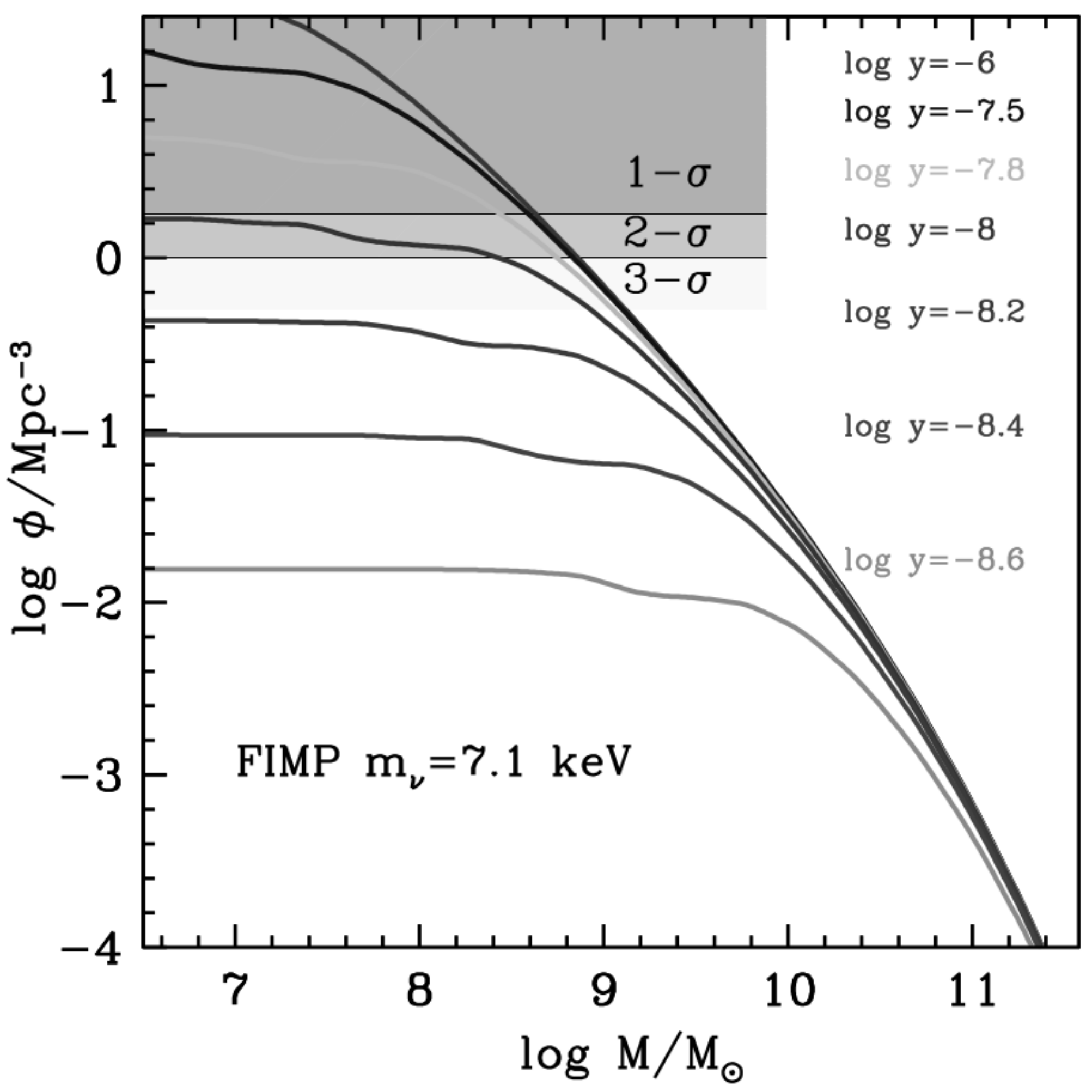}}\hspace{0.5cm}
        \caption{\it Adapted from \citet{Menci2017}: same as Fig.~\ref{fig1} for Resonant Production (left) and Scalar Decay with small Higgs portal coupling (right) sterile neutrino DM models. The illustrative case with $m_{\nu}=7.1$~keV, corresponding to a particle whose decay could be at the origin of the potential 3.5~keV line in X-ray spectra of clusters, is considered. The cumulative mass functions are derived at varying mixing amplitude (from $10^{-12}$ to $5\times10^{-10}$, bottom to top) and Yukawa coupling $y$ (log($y$) from -8.6 to -6, bottom to top) for the RP and SD case respectively.}
\label{fig2}
    \end{center}
\end{figure}

In the case of resonantly produced sterile neutrino DM, we choose the free parameters to be the mass, $m_{\nu}$, and the mixing amplitude $\sin^2(2\theta)$. We first investigate the effect of varying the mixing angle for a fixed sterile neutrino mass by focusing on the case $m_{\nu}=7.1$~keV, corresponding to a sterile neutrino whose decay could be at the origin of the potential 3.5~keV line in X-ray spectra of clusters. For such a case, the spectra yield the cumulative halo mass functions shown in Fig.~\ref{fig2} (left panel) for different values of $\sin^2(2\theta)$. The condition on the number density of DM halos to be larger than the observed abundance $\overline{\phi}\geq {\phi}_{obs}$ restricts the mixing angle in the range $2\times10^{-11}\leq \sin^2(2\theta)\leq 10^{-9}$ (at 2-$\sigma$ confidence level). 
 
\begin{figure}[htb]
    \begin{center}
     {\includegraphics[scale=0.4]{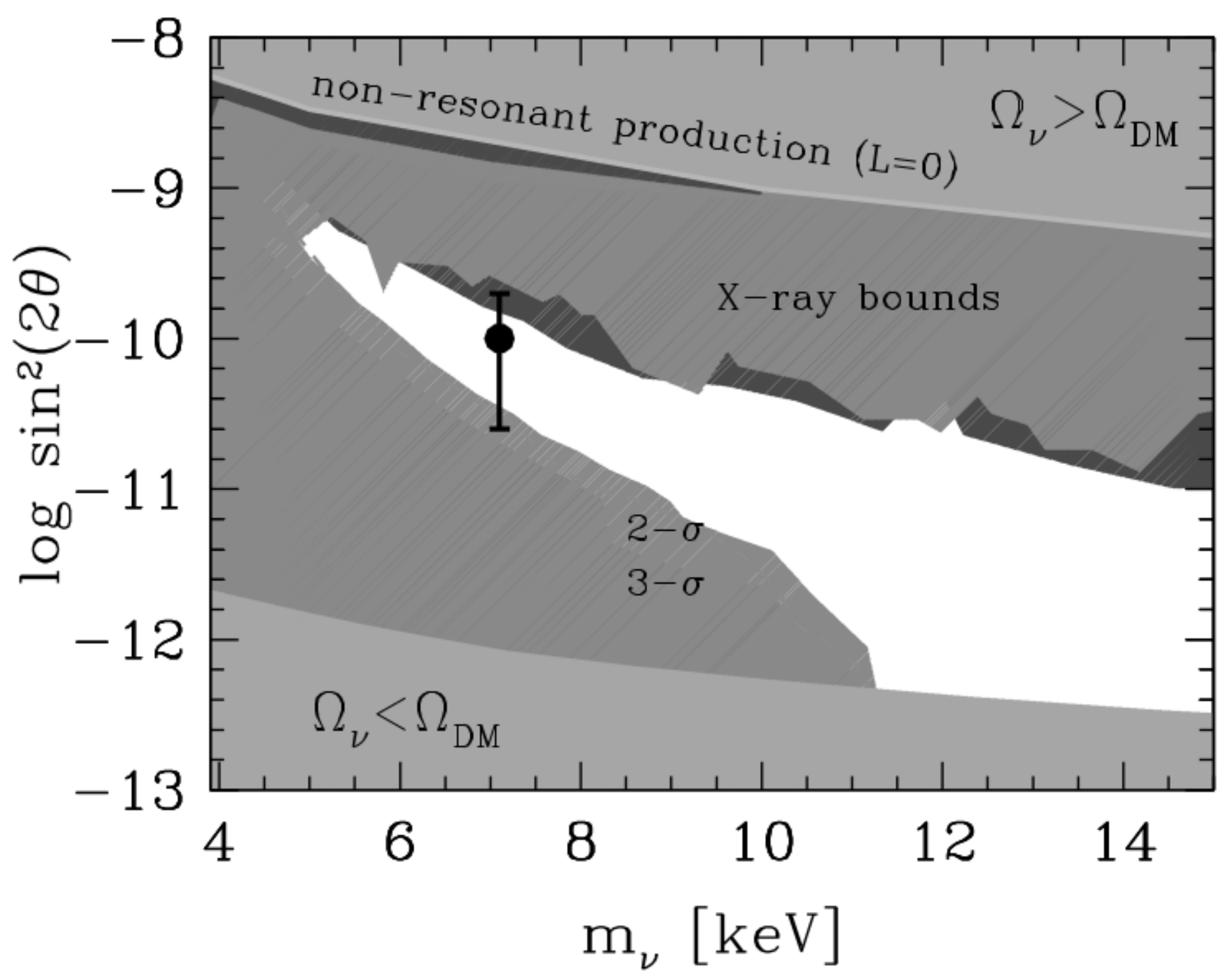}}\hspace{0.5cm}      
        \caption{\it Adapted from \citet{Menci2017}: the constraints on the RP sterile neutrino parameter space from our method are represented as exclusion regions,  with 3-$\sigma$ and 2-$\sigma$ limits represented by darker and lighter colors, together with other constraints from the literature \citep[see][for details]{Menci2017}. The tentative line signal at 7.1~keV is shown by the point with error bars. }
\label{fig3}
    \end{center}
\end{figure}

We also explore the whole range of free parameters using a grid of values for both $m_{\nu}$ and $\sin^2(2\theta)$. After computing the corresponding power spectra, the condition $\overline{\phi}\geq {\phi}_{obs}$ leads to the exclusion region in the plane $m_{\nu}-\sin^2(2\theta)$ shown in Fig.~\ref{fig3}. We exclude all models with a sterile neutrino mass below $m_{\nu}\leq 5$~keV and also large parts of the parameter space above. 

 \subsection{Sterile Neutrino from scalar decay}

In the case of SD sterile neutrinos the parameter space is three-dimensional, since it includes the mass of the scalar $m_S$, the Higgs portal coupling $\lambda$ and the Yukawa coupling with the scalar $y$. We show in Fig.~\ref{fig2} (right panel) a comparison between the model cumulative halo distributions and the observed number density of galaxies in the illustrative case of a sterile neutrino with $m_{\nu}=7.1$~keV (the candidate origin of the potential 3.5~keV line) in the limit of small Higgs portal coupling $\lambda\ll 10^{-6}$. In this case, the present data allow to set a constraint $y\geq 9\times10^{-9}$ at 2-$\sigma$ confidence level. 
 
We then extend our exploration to cover the whole parameter space of SD production model for sterile neutrinos. To this aim, we consider a grid of $\lambda$ and $y$ values for six different values of the scalar mass $m_S/{\rm~GeV}=60, 65, 100, 170, 500, 1000$. For each value of $m_S$, we compute the power spectrum corresponding to each point in the $\lambda-y$ plane. In Fig. 4 of \citet{Menci2017} we show the regions of the parameter space consistent with the galaxy number densities measured in the HFF ($\overline{\phi}\geq \phi_{\rm obs}$). These  regions clearly split into a freeze-out (for $\lambda\geq 10^{-6}$) and  freeze-in (for $\lambda\ll 10^{-6}$) family.  For the freeze-out family, decreasing the scalar mass  $m_S$ leads to a tighter bound on $y$, while yielding an approximate lower bound of $\lambda\gtrsim 10^{-5.2}$ for the Higgs portal coupling. For the freeze-in family, decreasing the scalar mass $m_S$ pushes the admitted values of $\lambda$ to progressively smaller values, while providing progressively stronger limits on $y$. 
\begin{figure}[htb]
    \begin{center}
        {\includegraphics[scale=0.35]{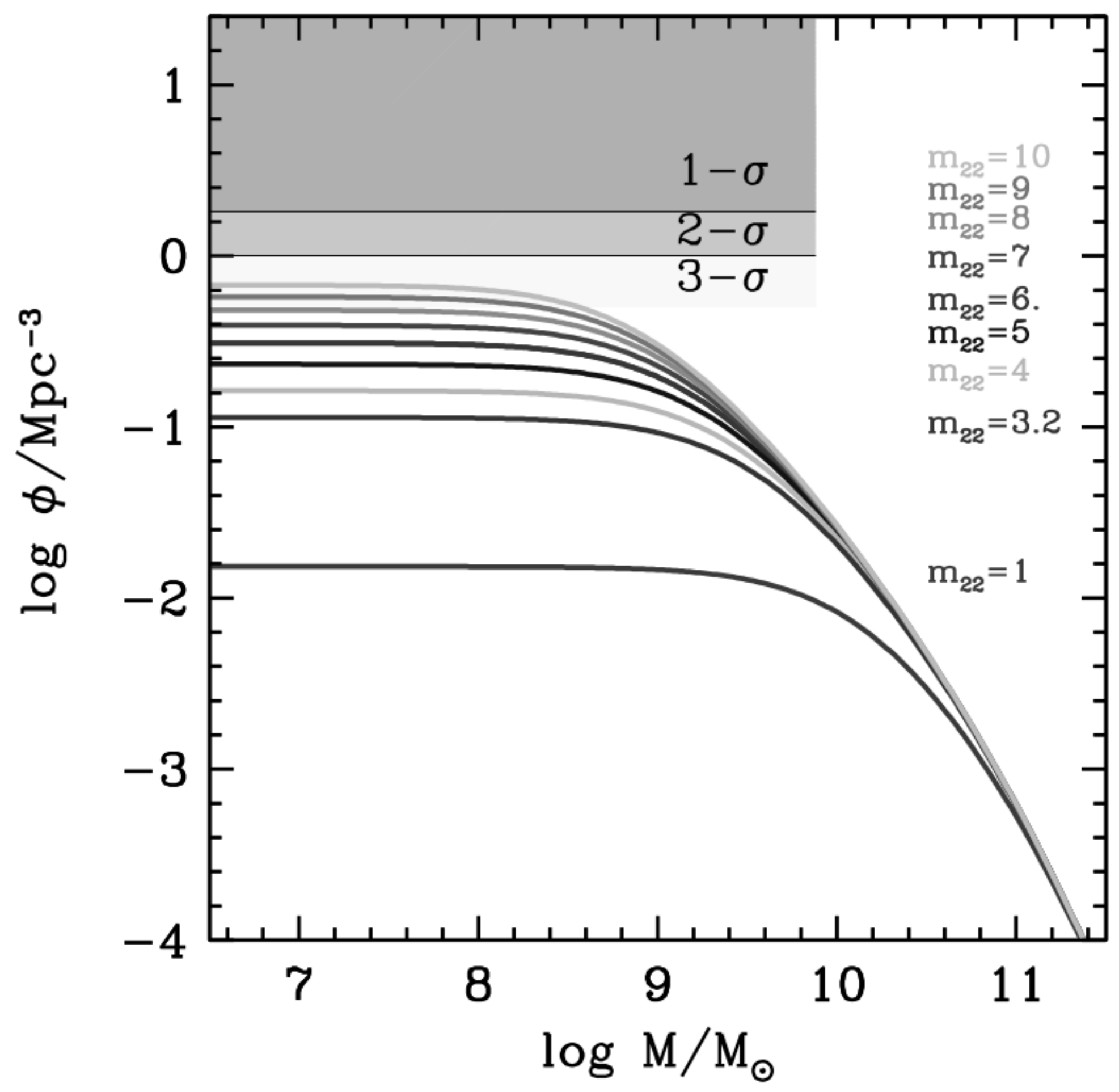}}\hspace{0.5cm}
        \caption{\it Adapted from \citet{Menci2017}: same as Fig.~\ref{fig1} for Fuzzy DM models with varying particle mass from 1 to 10 in units of $10^{-22}$~eV (bottom to top).}
\label{fig4}
    \end{center}
\end{figure}

 \subsection{Fuzzy DM}
 
The large observed number density  of high redshift galaxies turns out to provide particularly strong constraints on Fuzzy DM.  In  Fig.~\ref{fig4} we show the cumulative halo mass function for different values of the DM particle mass (in units of $10^{-22}$~eV). The strong suppression in the number of low-mass halos compared to the CDM case yields a lower limit $m_\psi\geq 10^{-21}$ eV for the DM particle mass at 3-$\sigma$ confidence level. Our results constitute the \emph{tightest constraint on Fuzzy DM particles derived so far}, and have a strong impact for the whole class of models based on Fuzzy DM. In fact, all results in the literature indicate that the mass of Fuzzy DM particles should be in the range $m_\psi=(1-5.6)\cdot 10^{-22}$~eV to explain the observed  density profile of nearby dwarf galaxies~\citep[e.g.,][]{gonzalez2016}. This is inconsistent at more than 3-$\sigma$ confidence level with our lower limits, strongly disfavoring such scenarios.

\section{Summary and conclusions}
The recently measured UV luminosity functions (LFs) of ultra-faint lensed galaxies at $z\approx 6$ provide 
strong constraints on DM models with suppressed power spectra. The comparison of the predicted maximum number density of DM halos $\overline{\phi}$ to the observed number density ${\phi}_{obs}$ provide robust constraints through the simple condition that observed galaxies cannot outnumber their host DM halos ($\overline{\phi}\geq {\phi}_{obs}$). Remarkably, these constraints are conservative, and independent of the modeling of baryonic physics in low-mass galaxies. The mass of WDM thermal relic candidates is constrained to be $m_X\geq 2.9$ keV at 1$\sigma$ confidence level,  and $m_X\geq 2.4$ keV at $2-\sigma$ level. The parameter space for RP and SD sterile neutrino models is significantly restricted. By taking the notable case of sterlie neutrinos whose decay can explain the potential 3.5~keV line ($m_{\nu}=7.1$~keV), the mixing amplitude in the RP case is restricted to  $-11.4\leq\log \sin^2(2\theta)\leq-10.2$, while the Yukawa coupling $y$ for SD production is constrained to $y \geq 9\times 10^{−9}$ at 2-$\sigma$ confidence level.

While our method is robust and independent of the baryon physics entering galaxy formation, we note that the measurements of the luminosity functions from strongly lensed galaxies are particulary delicate at the faint end where large magnifications are involved and where the computation of effective volumes is prone to subtle systematic effects. As an example, \citet{Bouwens2016} have adopted  a different estimate of the impact of lensing magnifiction finding not only a lower median value for the number density of galaxies  at $M_{UV}=-12.5$ compared to \cite{Livermore:2016mbs}, but also larger error bars, resulting in looser constraints on the parameters of DM models. A thorough discussion of the impact of observational uncertainties on our constraints is provided in \citet{Menci2016} and \citet{Menci2017}.

The main step to provide more stringent constraints is thus clearly a deeper understanding of the systematics associated with the lensing observations of faint, high-redshift galaxies. Refined lensing models and more accurate determinations of the source redshifts, together with the inclusion of observational data from other strong-lensing clusters will certainly enable an improved comparison between the observed galaxy number density and predicted halo number density in a variety of DM scenarios. In a few years from now a significant leap will be made possible by deep JWST imaging reaching absolute magnitudes of $M_{\rm UV}\approx -11$ on 5 times larger samples of high-redshift galaxies.

\end{document}